\documentclass[reprint,showpacs,twocolumn,jcp,nobibnotes]{revtex4-1}
\usepackage{hyperref}
\usepackage[dvipsnames]{xcolor}
\hypersetup{colorlinks=true, citecolor=blue, urlcolor=blue, linkcolor=blue}
\usepackage{graphics,graphicx,amsfonts,amsmath,amsbsy,amssymb,color}
\usepackage{bm}
\usepackage{subfigure}
\usepackage{threeparttable}
\usepackage{amsmath}
\usepackage{float}

\usepackage{booktabs}
\AtBeginDocument{
  \heavyrulewidth=.08em
  \lightrulewidth=.05em
  \cmidrulewidth=.03em
  \belowrulesep=.65ex
  \belowbottomsep=0pt
  \aboverulesep=.4ex
  \abovetopsep=0pt
  \cmidrulesep=\doublerulesep
  \cmidrulekern=.5em
  \defaultaddspace=.5em
}

\newcommand{\ket}[1]{\left|{#1}\right>}

\newcommand{\braket}[3]{\left \langle {#1}\middle|{#2}\middle|{#3}\right \rangle}
\newcommand{\brakets}[2]{\left<{#1}\middle|{#2}\right>}
\newcommand{\psipair}[1]{\Psi_{ij}^{#1}}
\newcommand{\dpair}[1]{\delta_{ij}^{#1}}
\newcommand{\RN}[1]{%
  \textup{\uppercase\expandafter{\romannumeral#1}}%
}
\newcommand{\dmp}{$\Delta E^\textrm{mp2}$}
\newcommand{\dppl}{$\Delta E^\textrm{ppl}$}
\newcommand{\dpsppl}{$\Delta E^\textrm{ps-ppl}$}
\newcommand{\drest}{$\Delta E^\textrm{rest}$}

\newcommand{\sdppl}{$\Delta \textrm{ps-ppl}$}

\begin{document}

\title{%
Focal-point approach with pair-specific cusp correction for coupled-cluster theory
}

\author{Andreas Irmler}
\email{andreas.irmler@tuwien.ac.at}
\author{Alejandro Gallo}
\author{Andreas Gr\"uneis}
\affiliation{
  Institute for Theoretical Physics, TU Wien,\\
  Wiedner Hauptstra{\ss}e 8--10/136, 1040 Vienna, Austria
}

\begin{abstract}

We present a basis set correction scheme for the coupled-cluster singles and
doubles (CCSD) method. The scheme is based on employing frozen natural orbitals (FNOs)
and diagrammatically decomposed contributions to the electronic correlation
energy that dominate the basis set incompleteness error (BSIE).  As recently
discussed in [\href{https://doi.org/10.1103/PhysRevLett.123.156401}{Phys. Rev.
Lett. 123, 156401 (2019)}], the BSIE of the CCSD correlation energy is
dominated by the second-order M{\o}ller-Plesset (MP2) perturbation energy and
the particle-particle ladder term.  Here, we derive a simple approximation to
the BSIE of the particle-particle ladder term that effectively corresponds to a
rescaled pair-specific MP2 BSIE, where the scaling factor depends on the
spatially averaged correlation hole depth of the
coupled-cluster and first-order pair wavefunctions.
The evaluation of the derived expressions is simple to implement in any existing code.
We demonstrate the effectiveness of the method for the uniform electron gas.
Furthermore, we apply the method to coupled-cluster theory calculations of atoms
and molecules using FNOs.
Employing the proposed correction and an increasing number of FNOs per occupied orbital,
we demonstrate for a test set that rapidly convergent
closed and open-shell reaction energies, atomization energies, electron affinities, and
ionization potentials can be obtained. 
Moreover, we show that a similarly excellent trade-off between required virtual orbital
basis set size and remaining BSIEs can be achieved for the
perturbative triples contribution to the CCSD(T) energy employing FNOs and the (T*) approximation.

\end{abstract} \date{\today}


\maketitle

\section{Introduction}

Traditional quantum chemical theories approximate the many-electron
wavefunction by a linear combination of Slater determinants constructed from
one-electron orbitals.  Unfortunately, this expansion causes a frustratingly
slow convergence of many calculated properties to the complete basis set limit.
This means that a significant part of the computational cost in many-electron
perturbation theory calculations originates from the need to include large
numbers of one-electron basis functions to achieve the desired level
of precision.  Many techniques have been developed to accelerate the
convergence to the complete basis set limit including explicitly correlated
methods, transcorrelated methods and focal-point
approaches~\cite{Kong:CR112-75,Hattig:CR112-4,Werner:JCP126-164102,Kutzelnigg:JCP94-1985,
Knizia2009,Tenno2017,Boys1969a,Boys1969b,Cohen2019, TENNO2000169, East1993, Csaszar1998,
Sinnokrot2002, Takatani2010, Jurecka2006, Warden2020, Carrell2012}.
More recently, a density-functional theory based approach has also been employed to correct
for basis set incompleteness errors~\cite{Giner2018}.
Furthermore, there exist a wide range of techniques
that aim at extrapolating to the complete basis set limit~\cite{Feller2013,Ranasinghe2013,Schwenke2005}.

Explicitly correlated methods are undeniably among the most reliable and
efficient methods to correct for the basis set incompleteness error (BSIE) in quantum
chemical many-electron theories. They account for the first-order cusp
condition in the many-electron wavefunction ansatz explicitly
and are commonly referred to as F12 theories, where F12 stands for a
two-electron correlation factor that enables a compact expansion of
the wavefunction at short interelectronic
distances~\cite{Kong:CR112-75,Hattig:CR112-4,Werner:JCP126-164102,Kutzelnigg:JCP94-1985,Knizia2009,Tenno2017}.
Despite the need for additional many-electron integrals such as three- and
sometimes even four-electron integrals, explicitly correlated coupled-cluster
theories have been implemented in a computationally efficient manner, allowing
for numerically stable and precise {\it ab initio} calculations of
molecular systems~\cite{Knizia2009,Hattig2010-ql,
Hattig:CR112-4,Kong:CR112-75,Tew:BOOK2010,
Valeev2012,Werner:BOOK2010,Tenno12,Tenno12_2,Shiozaki2010}. 

Focal-point approaches seek to combine low and high-level theories in a
computationally efficient manner to correct for the BSIE in the high-level
calculation. In this work we seek to correct for the BSIE of coupled-cluster
theory using corrections based on second-order M\o ller-Plesset perturbation
theory (MP2).  Unfortunately, the BSIEs of MP2 and coupled-cluster singles and
doubles (CCSD) energies usually differ significantly.  This can be attributed
to the so-called ``interference effect'' that leads to a reduction in the BSIE
of CCSD theory compared to MP2 theory~\cite{Nyden1981,Petersson1981}.  Peterson
{\it et al.} account for this effect using complete basis set (CBS) limit
model chemistry via an empirical formula~\cite{Petersson1988}.  Similar
approaches include rescaled explicitly correlated MP2
energies~\cite{Vogiatzis2011}.  We stress that the particle-particle ladder
(PPL) contribution to the CCSD correlation energy is the leading cause of the
interference effect.  Therefore, its contribution to the BSIE needs to be
treated as accurately as possible to attain well converged CCSD
energies~\cite{Valeev2007,Irmler2019a,Irmler2019b}.  In this context, it is
noteworthy that studies on the basis set convergence in third-order
perturbation theory have also revealed the significance of corresponding PPL
contributions~\cite{Kutzelnigg92,Ohnishi2013}.

Recently, we have explored a focal-point approach to correct for the BSIE of CCSD
correlation energies that is based on its diagrammatic decomposition into
an MP2, PPL and so-called rest term~\cite{Irmler2019a}.
The decomposition of the CCSD correlation energy into contributions that dominate
the BSIE (MP2 and PPL) and the rest can be very useful in {\it ab initio}
calculations for the following two reasons. Firstly, the MP2 correlation energy in the
complete basis set limit can be computed using algorithms with low
computational complexity, removing the corresponding BSIE almost
exactly~\cite{Schuetz1999,Schaefer2017,Bircher2020,Neuhauser2013}.
Secondly, constructing approximations to
the BSIE in the PPL term is a much simpler challenge than correcting the BSIE
in all possible diagrammatic contributions to the CCSD correlation energy.
Furthermore, the MP2 and PPL terms can be regarded as an independent electron
pair approximation to CCSD theory, which simplifies the development of an electron
pair-specific BSIE correction.  In Ref.~\cite{Irmler2019b} we have shown that
even a simple correction to the PPL BSIE, that is based on rescaling a
corresponding MP2 correction, achieves significant reductions in the BSIE of
the CCSD correlation energy. However, the attained level of precision
did not reach F12-like quality for all investigated properties.

Here, we present a computationally efficient and significantly more
accurate method to correct for the BSIE in the PPL term of the CCSD correlation
energy.  Compared to our previous work in Ref.~\cite{Irmler2019b}, the
improvements result from the following two modifications. Firstly, we employ frozen
natural orbitals for the virtual orbital manifold
calculated from approximate one-particle reduced density matrices~\cite{Taube2005}.
Secondly, and most importantly, we introduce approximations
to the electron pair wavefunctions appearing in the expression for the PPL term,
making it possible to compute BSIE corrections with low computational cost but
high accuracy.  To this end, we develop a pair-specific mean-field ansatz that
exhibits an average correlation hole depth that agrees with the spatially
averaged correlation hole depth of the coupled-cluster or first-order wavefunction in
a finite basis set.  Employing this mean-field ansatz in the bra or
ket state of the PPL expression leads to considerable simplifications.
The derived pair-specific
BSIE correction effectively corresponds to the rescaled MP2 BSIE.
As already mentioned above, the scaling factor depends on averaged
correlation hole depths.
In hindsight, our work corroborates the success of CCSD basis set corrections
that are based on rescaling MP2 BSIEs.
We note that the averaged pair-specific correlation hole depth can be computed
from the electronic transition structure factor, which introduces the need for
two-electron integrals employing the $\delta({\mathbf r}_{12})$-kernel.
However, these additional integrals do not increase significantly the computational
complexity of a conventional MP2 calculation in a finite basis
set.  We demonstrate that the introduced BSIE correction yields CCSD
correlation energies that converge rapidly with respect to the number of frozen
natural orbitals.  





%
\section{Theory}
\label{sec:theory}

\subsection{The pair-specific decomposed PPL correlation energy in the CBS limit}
In this work we introduce a correction to the BSIE of the CCSD correlation energy.
The CCSD correlation energy is given by
\begin{equation}
E^{\rm CCSD}_{\rm c}= \sum_{ij} \sum_{ab}
	T_{ij}^{ab} \left( 2 \braket{ij}{V}{ab} - \braket{ji}{V}{ab} \right)
\label{eq:Ecorr3}
\end{equation}
and can be decomposed into different diagrammatic contributions such that~\cite{Irmler2019a}
\begin{equation}
E_{\rm c}^{\rm CCSD}=E^{\rm mp2}+  E^{\rm ppl}+ \underbrace{ E^{\rm phl}+ E^{\rm hhl}+ E^{\rm phr}+ \ldots }_{E^{\rm rest}},
\label{eq:Ecorr}
\end{equation}
where $E^{\rm mp2}$ corresponds to the MP2 correlation energy
\begin{equation}
E^{\rm mp2}= \sum_{ij} \sum_{ab} W_{ab}^{ij} \braket{ab}{V}{ij}
\label{eq:Ecorr2}
\end{equation}
and the particle-particle ladder term is defined as
\begin{equation}
E^{\rm ppl}= \sum_{ij} \sum_{ab}
W_{ab}^{ij} \sum_{cd} \braket{ab}{V}{cd} T_{ij}^{cd}.
\label{eq:EcorrPPL}
\end{equation}
We note that \( E^{\mathrm{mp2}} \) was referred to as \( E^{\mathrm{driver}}\)
in previous related work \cite{Irmler2019a,Shepherd2014}.  $T_{ij}^{cd}$ is
computed from the CCSD singles ($t_i^a$) and doubles ($t_{ij}^{ab}$) amplitudes
as $T_{ij}^{ab}=t_{ij}^{ab}+t_i^a t_j^b$.  $t_i^a$ and $t_{ij}^{ab}$ are
obtained by solving the corresponding amplitude
equations\cite{Bartlett07,Cizek71}.  $W_{ab}^{ij}$ is given by
\begin{equation}
W_{ab}^{ij}= \frac{
2 \braket{ij}{V}{ab}- \braket{ji}{V}{ab} }
{\epsilon_i + \epsilon_j - \epsilon_a - \epsilon_b}.
\label{eq:Wijab}
\end{equation}
All equations refer to spin-restricted spatial orbitals with the Coulomb integrals defined by
\begin{equation}
\label{eqn:integral}
\braket{ij}{V}{ab} = \iint d\mathbf{r}_1 d\mathbf{r}_2 
     \phi^*_i (\mathbf{r}_1) \phi^*_j (\mathbf{r}_2) v(r_{12})
     \phi_a (\mathbf{r}_1) \phi_b (\mathbf{r}_2) \text,
\end{equation} 
using the Coulomb kernel $v(r_{12}) = 1/\left|\mathbf{r}_1-\mathbf{r}_2\right|$.

The following discussion is based on the premise that the finite virtual orbital
manifold is spanned by a set of canonical orbitals that needs to be augmented
with additional virtual orbitals to reach the CBS limit, while the occupied
orbitals are fully converged to the CBS limit regardless of the approximations
used in the virtual orbital space.
This situation closely resembles {\em ab initio} calculations employing
re-canonicalized frozen natural orbitals~\cite{Taube2005}.
We choose the following index labels for occupied and virtual spatial orbitals
\begin{center}
\begin{tabular}{ l l }
 $i$, $j$, $k$, $\hdots$  & occupied states  \\
 $a$, $b$, $c$, $\hdots$  & virtual states in finite basis \\
 $\alpha$, $\beta$, $\gamma$, $\hdots$ & augmented virtual states\\
 $A$, $B$, $C$, $\hdots$  & union of all virtual states. \\
\end{tabular}
\end{center}

%
%

The particle-particle ladder term ($E^{\rm ppl}$) defined by Eq.~(\ref{eq:EcorrPPL})
can also be expressed as
\begin{equation}
\epsilon^\textrm{A}_{ij} = \left<\Psi_{ij}^{(1)} \middle| V \middle|
              \Psi_{ij}^{\textrm{cc}} \right> \text,
\label{eq:EcorrPPLPsi}
\end{equation}
where
\begin{equation}
\left< \psipair{\textrm{(1)}} \right| = \sum_{ab} 
                 W_{ab}^{ij} \left< ab \right|,
\label{eqn:psiccdef}
\end{equation}
and
\begin{equation}
\left| \psipair{\textrm{cc}} \right> = \sum_{cd} T_{ij}^{cd} \left| cd \right>.
\label{eqn:psiccdef}
\end{equation}
Herein
\( \left|\psipair{\textrm{(1)}} \right\rangle \)
refers to the linearized first
order wavefunction whereas
\(\left| \psipair{\textrm{cc}} \right>\)
resembles a coupled-cluster pair wavefunction.
Consequently, the PPL term only couples wavefunctions with the same occupied pair index.


We formally define the CBS limit of the linearized first-order
and coupled-cluster-like pair wavefunction by
\begin{equation}
\ket{\Psi_{ij}^{(1)\textrm{-cbs}}} = \ket{\Psi_{ij}^{(1)}} + \ket{\delta_{ij}^{(1)}}
\end{equation}
and
\begin{equation}
\ket{\Psi_{ij}^{\textrm{cc-cbs}}} = \ket{\Psi_{ij}^{\textrm{cc}}} + \ket{\delta_{ij}^{\textrm{cc}}},
\end{equation}
respectively. $\ket{\delta_{ij}^{(1)}}$ and $\ket{\delta_{ij}^{\textrm{cc}}}$
are defined such that they correct for the BSIE in the respective parent
wavefunctions $\ket{\Psi_{ij}^{(1)}}$ and $\ket{\Psi_{ij}^{\textrm{cc}}}$. Already
in 1985, Kutzelnigg discussed that the conventional expansion, using products
of one-electron states, can not represent the wavefunction accurately at regions where
the interelectronic distance approaches zero~\cite{Kutzelnigg1985}.
Thus, for increasing one-electron basis set sizes, the contribution of
$\ket{\delta_{ij}^{(1)}}$ and $\ket{\delta_{ij}^{\textrm{cc}}} $ will largely
be localized to the cusp region at small interelectronic distances.
Substituting the above BSIE corrections into Eq.~(\ref{eq:EcorrPPLPsi}), yields
the following contributions to the PPL energy in the CBS limit:
\begin{equation}
\label{eqn:epsB}
\epsilon^\textrm{B}_{ij} = \braket{\dpair{(1)}}{V}{\psipair{\textrm{cc}}}
\end{equation}
\begin{equation}
\label{eqn:epsC}
\epsilon^\textrm{C}_{ij} = \braket{\psipair{(1)}}{V}{\dpair{\textrm{cc}}}
\end{equation}
\begin{equation}
\label{eqn:epsD}
\epsilon^\textrm{D}_{ij} = \braket{\dpair{(1)}}{V}{\dpair{\textrm{cc}}}.
\end{equation}
Consequently, the CBS limit formally reads
\begin{equation}
	E^\textrm{ppl-cbs} = \sum_{ij} \left ( \epsilon^\textrm{A}_{ij} + \epsilon^\textrm{B}_{ij}
		   + \epsilon^\textrm{C}_{ij} + \epsilon^\textrm{D}_{ij} \right )
                   .
\end{equation}
This work outlines an efficient approximation
to \( E^{\mathrm{ppl-cbs}} \).
To this end, we analyze
$\epsilon^\textrm{B}_{ij}$ and $\epsilon^\textrm{C}_{ij}$
and provide suitable approximations for them in the
following sections.
We disregard
$\epsilon^\textrm{D}_{ij}$
since it is of second order in the BSIE of the pair wavefunctions
($\delta$). We recall that $\epsilon^\textrm{A}_{ij}$ is evaluated
in the conventional coupled-cluster calculation using the finite basis set.

\subsection{Coupling of $\ket{\delta_{ij}^{(1)}}$ to $\ket{\Psi_{ij}^{\mathrm{cc}}}$}
We now turn to the expression for
$\epsilon^\textrm{B}_{ij}$ and employ the resolution of the identity (RI)
\begin{equation}
\label{eqn:epsb}
\begin{split}
	\epsilon^\textrm{B}_{ij} = &
  \sum_{\alpha \beta} \brakets{\dpair{(1)}}{\alpha \beta} \braket{\alpha \beta}{V}{\psipair{\textrm{cc}}} \\
	& + \sum_{a\beta}\brakets{\dpair{(1)}}{a \beta} \braket{a \beta}{V}{\psipair{\textrm{cc}}} \\
	& + \sum_{\alpha b}\brakets{\dpair{(1)}}{\alpha b} \braket{\alpha b}{V}{\psipair{\textrm{cc}}} .
\end{split}
\end{equation}
The above equation can be interpreted as the coupling of the change of the
first-order wavefunction to $|\psipair{\textrm{cc}}\rangle$.
Due to $ \brakets{\dpair{(1)}}{a b} = 0$, only projectors that involve at least
one state from the augmented virtual basis have to be included in the
RI.
The orbitals $\phi_\alpha$, and $\phi_\beta$
are strongly oscillating in space, {\it i.e.}, they bear large wave number
and/or high angular
momentum number. In contrast, $|\psipair{\textrm{cc}}\rangle$ is expected to be
much smoother. Following fundamental ideas of scattering theory, we
replace the complicated scattering problem with a much simpler one, by means of the
following approximation
%
\begin{equation}
\left| \psipair{\textrm{cc}} \right> =
\sum_{cd} T_{ij}^{cd} \left| cd \right> \approx \left| ij \right> g^{\textrm{cc}}_{ij},
\label{eqn:psicc}
\end{equation}
%
%
%
where $\left| ij \right>$ is a mean-field state constructed
from the Hartree--Fock orbitals of the occupied pair $i$ and $j$.
The scaling factor $g^{\textrm{cc}}_{ij}$ is chosen such that the spatially averaged
correlation hole depths of the correlated wavefunction and
its mean-field approximation are equated after projection onto the
occupied space of the same electron pair:
\begin{equation}
	\sum_{cd} T_{ij}^{cd} \left< ij | \delta(\mathbf{r}_{12})  | cd \right>
	=  \left< ij | \delta(\mathbf{r}_{12}) | ij \right> g^{\textrm{cc}}_{ij}.
\label{eqn:matching}
\end{equation}
The appearing integrals are defined in an analogous manner to Eq.~(\ref{eqn:integral}) but
with the Coulomb kernel replaced by the Dirac delta function $\delta(\mathbf{r}_{12})$.
When using Gaussian basis functions, this requires only minor modifications
of the original integral routines (see \cite{Ahlrichs2006}).
From the above equation, we obtain an explicit expression for the
pair-specific correlation hole depth scaling factor given by
\begin{equation}
	g^{\textrm{cc}}_{ij} = \frac{ \sum_{cd} T_{ij}^{cd} \left< ij | \delta(\mathbf{r}_{12})  | cd \right> } { \left< ij | \delta(\mathbf{r}_{12}) | ij \right> }.
\label{eqn:gij}
\end{equation}
For the sake of brevity in the following paragraphs, we introduce a
projection operator $\hat{g}_{ij}$ that
yields an approximate mean-field state when applied to any
correlated electron pair state, such that
\begin{equation}
\label{eqn:mf}
\left| ij \right> g^{\textrm{cc}}_{ij} = \hat{g}_{ij} \left| \psipair{\textrm{cc}} \right>.
\end{equation}

To get a better understanding of the above approximation, we now inspect the
explicit expression for $\epsilon^\textrm{B}$ of a singlet state, which is
given by
\begin{equation}
	\braket{\delta}{V}{\psi}
	= \iint d\mathbf{r}_{12} d\mathbf{\bar{r}}_{12} {\tilde{\delta}}^*(\mathbf{r}_{12},\bar{\mathbf{r}}_{12}) \frac{1}{|\mathbf{r}_{12}|} \tilde{\psi}(\mathbf{r}_{12},\bar{\mathbf{r}}_{12}).
\end{equation}
Here, the electron-pair functions $\delta$ and $\psi$ have been transformed
to a real-space representation in $\mathbf{r}_{12}={\mathbf{r}_1-\mathbf{r}_2}$
and $\bar{\mathbf{r}}_{12}={\mathbf{r}_1+\mathbf{r}_2}$.
Because $\tilde{\delta}$ is largely localized around the cusp region, it effectively screens the
Coulomb kernel at large interelectronic distances $|\mathbf{r}_{12}|$.
$\epsilon^\textrm{B}$ is therefore dominated by contributions from short interelectronic
distances. Moreover, $\tilde{\psi}$ is a smooth function in the cusp region compared
to $\tilde{\delta}$, suggesting that
\begin{equation}
\tilde{\psi}(\mathbf{r}_{12},\bar{\mathbf{r}}_{12}) \approx  \tilde{\psi}(\mathbf{r}_{12}=0,\bar{\mathbf{r}}_{12})
\end{equation}
is a reasonable approximation.
$\tilde{\psi}(0,\bar{\mathbf{r}}_{12})$ is the correlation hole depth as a
function of $\bar{\mathbf{r}}_{12}$.  
The central approximation of this work is based on employing a mean field ansatz for
$\tilde{\psi}(0,\bar{\mathbf{r}}_{12})$ that is obtained by
projecting $\tilde{\psi}(0,\bar{\mathbf{r}}_{12})$ onto a corresponding
zeroth-order mean-field wave function and ensuring that
the spatially averaged correlation depths of the mean-field ansatz and $\tilde{\psi}$ agree.
This is achieved using the pair-specific projection operator $\hat{g}_{ij}$ defined
in Eq.~(\ref{eqn:mf}).

Using the mean-field approximation described above, 
$\epsilon_{ij}^\textrm{B}$ can be approximated as follows
\begin{equation}
\label{eqn:epsbapprox}
\begin{split}
\epsilon_{ij}^\textrm{B} = \braket{\dpair{(1)}}{V}{\psipair{\textrm{cc}}}  &\approx
	\underbrace{ \braket{\dpair{(1)}}{V}{ij} }_{\Delta \epsilon_{ij}^{(2)}}  g^\textrm{cc}_{ij} 
\end{split}
\end{equation}
where $\Delta \epsilon_{ij}^{(2)}$ refers to the pair-specific
BSIE correction of the MP2 correlation energy.
Thus, we have shown that the $\epsilon_{ij}^\textrm{B}$
contribution to the PPL term can be approximated using $\Delta \epsilon_{ij}^{(2)}$ times a scaling factor
that depends on the spatially averaged correlation hole
depth of $\left | \psipair{\textrm{cc}} \right>$.

\subsection{Coupling of $\ket{\delta_{ij}^{cc}}$ to $\ket{\Psi_{ij}^{\mathrm{(1)}}}$}
We now focus on the coupling between the first-order wavefunction and
the BSIE correction to $\psipair{\textrm{cc}}$.
Using once more the RI we write Eq.~(\ref{eqn:epsC}) in the following way
\begin{equation}
\label{eqn:epsc}
\begin{split}
\epsilon^\textrm{C}_{ij} &=
 \sum_{CD}  \braket{\psipair{(1)}}{V}{CD} \brakets{CD}{\dpair{\textrm{cc}}} \\
&\approx \sum_{CD} \braket{\psipair{(1)}}{\hat{g}_{ij}^\dagger V}{CD} \brakets{CD}{\dpair{\textrm{cc}}} \\
&= g_{ij}^{(1)} \sum_{CD} \braket{ij}{V}{CD} \brakets{CD}{\dpair{\textrm{cc}}} \text.
\end{split}
\end{equation}
In the above equation, we have approximated the
first-order state by a mean-field state
that exhibits an identical spatially averaged correlation
hole depth.
Furthermore, the exact expression for
$\left| \delta_{ij}^\textrm{cc} \right>$ is not accessible, as we do not intend to solve the
coupled-cluster equations in the large basis set.
Moreover, we note that $\langle cd | \delta_{ij}^\textrm{cc} \rangle \neq 0$,
which is in contrast to the BSIE of the first-order state, where
$\langle cd | \delta_{ij}^\textrm{(1)} \rangle = 0$.
Therefore, we approximate the orbital representation of $\left| \delta_{ij}^\textrm{cc} \right>$ including only
the dominant contributions (driver and PPL) in the complete basis set limit of the amplitude equations:
\begin{equation}
\label{eqn:bone}
\begin{split}
	\brakets{CD}{\dpair{\textrm{cc}}} \approx  & \underbrace { \brakets{CD}{\dpair{(1)}} }_{ (\RN{1}) } \oplus
	    \underbrace{ \ \frac{ \braket{\gamma \zeta}{V}{\psipair{\textrm{cc}}}  }{\epsilon_{i}+\epsilon_j-\epsilon_\gamma -\epsilon_\zeta} }_{ (\RN{2}) }  \\
	& \oplus \underbrace { \ \frac{ \braket{C \zeta}{V}{\psipair{\textrm{cc}}}  }{ \epsilon_{i}+\epsilon_j-\epsilon_C -\epsilon_\zeta }  }_{ (\RN{3}) }
	\oplus \underbrace { \ \frac{ \braket{\gamma D}{V}{\psipair{\textrm{cc}}}  }{ \epsilon_{i}+\epsilon_j-\epsilon_\gamma -\epsilon_D } }_{ (\RN{4}) }  \\
        & \oplus \frac{ \braket{CD}{V}{\dpair{\textrm{cc}}}  }{\epsilon_i+\epsilon_j-\epsilon_C -\epsilon_D } \oplus \cdots
\end{split}
\end{equation}
The direct sum notation is used to emphasize the fact that $\gamma$ is a subset of $C$.
In the following we consider only those terms defined by $(\RN{1})-(\RN{4})$
because they are of zeroth-order in $\delta V$,
while the rest is $\mathcal{O}(\delta V)$.
We now turn to the contributions of the terms defined by $(\RN{1})-(\RN{4})$
to $\epsilon^{{\textrm C}}_{ij}$.
Inserting (\RN{1}) from Eq.~(\ref{eqn:bone}) into the last line of
Eq.~(\ref{eqn:epsc}) yields
\begin{equation}
	\epsilon^{\textrm{C}}_{ij}(\RN{1})=
	g_{ij}^{(1)} \Delta \epsilon_{ij}^{(2)}\textrm{.}
\end{equation}
To account for the contributions of $(\RN{2})-(\RN{4})$ to $\epsilon^{{\textrm C}}_{ij}$,
we again approximate $\left | \psipair{\textrm{cc}} \right>$ using the mean-field
ansatz defined by Eq.~(\ref{eqn:mf}).
Inserting the resulting approximations into the last line of Eq.~(\ref{eqn:epsc}) yields
\begin{equation}
	\epsilon^{\textrm C}_{ij}(\RN{2}-\RN{4})=
	g_{ij}^{(1)} \Delta \epsilon_{ij}^{(2)} g_{ij}^{\textrm{cc}}\textrm{.}
\end{equation}
Therefore, our final approximation to Eq.~(\ref{eqn:epsC}) is given by
\begin{equation}
\label{eqn:epscapprox}
	\epsilon^{\textrm C}_{ij}\approx\epsilon^{\textrm{C}}_{ij}(\RN{1})
	+\epsilon^{\textrm C}_{ij}(\RN{2}-\RN{4})=
	\Delta \epsilon_{ij}^{(2)} (
	g_{ij}^{(1)} +
	g_{ij}^{(1)} g_{ij}^{\textrm{cc}} ),
\end{equation}
which corresponds again to $\Delta \epsilon_{ij}^{(2)}$ scaled by a factor
that depends on the correlation hole depths of $\left | \psipair{\textrm{cc}} \right>$
and $\left | \psipair{\textrm{(1)}} \right>$.

We note that a corresponding BSIE correction in MP3 theory would have to
include $\epsilon^{\textrm{C}}_{ij}(\RN{1})$ only. 
However, in CCSD theory the BSIE
of $|\psipair{\textrm{cc}}\rangle$ is not well approximated using $\dpair{(1)}$.
Therefore $\epsilon^{\textrm C}_{ij}(\RN{2}-\RN{4})$ accounts for the change of
$\dpair{(1)}$ due to the most important PPL coupling terms linear in
$|\psipair{\textrm{cc}}\rangle$.  The coupling strength of these terms is on the order
of $g_{ij}^{\textrm{cc}}$ and needs to be included to attain high
accuracy.

\subsection{The pair-specific PPL basis-set correction}
We now summarize the final approximation to the BSIE correction of the PPL energy:
\begin{equation}
	\epsilon^{\textrm B}_{ij} + \epsilon^{\textrm C}_{ij}
	\approx
\Delta \epsilon_{ij}^{(2)} (
	g_{ij}^{\textrm{cc}} +
	g_{ij}^{(1)} +
	g_{ij}^{(1)} g_{ij}^{\textrm{cc}} ).
\end{equation}
We stress that the contribution of $\epsilon^{{\textrm D}}_{ij}$ defined in
Eq.~(\ref{eqn:epsD}) has been neglected because it is not of leading order in
$\delta$. We arrive at the following approximate CBS limit
expression of the PPL energy:
\begin{equation}
\label{eqn:Epsppl}
  E^\textrm{ps-ppl}=E^\textrm{ppl}+
  \underbrace{ \sum_{ij}\Delta \epsilon_{ij}^{(2)} (
        g_{ij}^{\textrm{cc}} +
        g_{ij}^{(1)} +
  g_{ij}^{(1)} g_{ij}^{\textrm{cc}} ) }_{\Delta \textrm{ps-ppl}}.
\end{equation}
At this point, we note again that in the above expression the pair-specific correlation
hole depth scaling factors $g_{ij}^{\textrm{cc}}$ and $g_{ij}^{(1)}$ are computed in a
finite basis set, whereas $\Delta \epsilon_{ij}^{(2)}$ refers to the BSIE correction
of the pair-specific MP2 correlation energy.

\section{The uniform two electron gas}
In order to assess the presented approximations, we first study a particularly
simple model system - the three-dimensional uniform electron gas (UEG).  The
details of this model are described for instance in Ref.~\cite{Shepherd2014}.
For the here performed analysis it is enough to study only two electrons in a
homogeneous positive background. The singlet ground state Hartree--Fock (HF)
wave function is a constant function and the virtual HF states are plane-waves
\begin{equation}
\phi_a (\mathbf{r}) = \frac{1}{\sqrt{\Omega}} e^{i\mathbf{k}_a\mathbf{r}} \text,
\end{equation}
with HF eigenvalues
\begin{equation}
	\epsilon_a = \frac{1}{2} {\mathbf{k}_a}^2 - \frac{4\pi}{\Omega\mathbf{k}_a^2} \text.
\end{equation}
Notice that the eigenenergies are ordered with respect to the length of the
corresponding  momentum vector. The unit cell volume is given by $\Omega$.
This simplifies the four-index integrals to
\begin{equation}
\braket{ii}{V}{ab} = \frac{4\pi}{\Omega \left|\mathbf{k}_a\right|^2}
              \delta_{\mathbf{k}_a+\mathbf{k}_b,0} \text{.}
\end{equation}
Consequently, the MP2 energy expression contains only a single sum 
\begin{equation}
	E^{\text{mp2}} = \sum_a \frac{\braket{ii}{V}{a\bar{a}}}{\epsilon_i+\epsilon_i-\epsilon_a-\epsilon_{\bar{a}}}
	\braket{a\bar{a}}{V}{ii} \text{.}
\end{equation}
%
We use the notation $\bar{a}$ for the
virtual orbital with momentum vector $-\mathbf{k}_a$.
The PPL energy expression reads
\begin{equation}
  E^{\text{ppl}}
  = \sum_a
\frac{ \braket{ii}{V}{a\bar{a}} }
       {\epsilon_i+\epsilon_i-\epsilon_a-\epsilon_{\bar{a}}}
	       \sum_c \braket{a\bar{a}}{V}{c\bar{c}} t_{ii}^{c\bar{c}}
\text{.}
\end{equation}
%

\begin{table*}
\small
\caption{
BSIEs for the two-electron UEG with $r_s = 3.5~a.u.$. Reference energies
are obtained from a calculation with 30046 virtuals. Referred to exact (ex.) is
the evaluation of Eq.~\eqref{eqn:uegabcd} for the converged amplitudes using
30046 virtuals and using $N_v$ orbitals in the finite basis. Estimates (est.)
are evaluated using Eqs.~\eqref{eqn:epsbapprox} and \eqref{eqn:epscapprox}.
The BSIE of the rest term is given in the last column and calculated between
results obtained with $N_v$ and 30046 virtual orbitals.
All energies are given in mH.
}
\label{tab:uegbcd}

\begin{tabular*}{0.6\textwidth}{@{\extracolsep{\fill}}lcccccccccc}
\toprule
$N_v$ &&
\multicolumn{2}{c}{$\epsilon^\textrm{B}$} &&
\multicolumn{2}{c}{$\epsilon^\textrm{C}$} &&
$\epsilon^\textrm{D}$ &&
$\Delta E^\textrm{rest}$
\\
&&
ex. &
est. &&
ex. &
est. &&
ex. &&
\\
\midrule
%
26  && 0.582 & 0.560 && 0.255 & 0.262 && 0.178 && -0.065 \\ 
56  && 0.343 & 0.332 && 0.157 & 0.162 && 0.076 && -0.049 \\ 
122 && 0.171 & 0.166 && 0.079 & 0.083 && 0.027 && -0.029 \\ 
250 && 0.087 & 0.084 && 0.047 & 0.043 && 0.010 && -0.015 \\ 
514 && 0.043 & 0.043 && 0.021 & 0.022 && 0.004 && -0.008 \\ 
\bottomrule
\end{tabular*}
\end{table*}

In a finite basis set calculation the number of virtual basis functions $N_v$
has to be truncated. For the UEG model system, this is typically done by
introducing a cutoff wave vector $k_1$ and considering only virtual states with
$|\mathbf{k}_a|<k_1$.  Following the ideas of section \ref{sec:theory}, we
introduce a second cutoff $k_2$ specifying the augmented virtual states
$\alpha$, with $k_1 \leq |k_\alpha| < k_2$.  Hence, we can write the following
four contributions to the total PPL energy

\begin{equation}
\label{eqn:uegabcd}
\begin{split}
	\epsilon_{ii}^\textrm{A} &= \sum_a
	\frac{ \braket{ii}{V}{a\bar{a}}  }{\epsilon_i+\epsilon_i-\epsilon_a-\epsilon_{\bar{a}}}
                        \sum_c \braket{a\bar{a}}{V}{c\bar{c}} t_{ii}^{c\bar{c}} \\
	\epsilon_{ii}^\textrm{B} &= \sum_\alpha
	\frac{ \braket{ii}{V}{\alpha\bar{\alpha}} }{\epsilon_i+\epsilon_i-\epsilon_\alpha-\epsilon_{\bar{\alpha}}}
                        \sum_c 
			\braket{\alpha\bar{\alpha}}{V}{c\bar{c}} t_{ii}^{c\bar{c}} \\
	\epsilon_{ii}^\textrm{C} &= \sum_a
	\frac{  \braket{ii}{V}{a\bar{a}} }{\epsilon_i+\epsilon_i-\epsilon_a-\epsilon_{\bar{a}}}
                        \sum_\gamma \braket{a\bar{a}}{V}{\gamma\bar{\gamma}} t_{ii}^{\gamma\bar{\gamma}} \\
	\epsilon_{ii}^\textrm{D} &= \sum_\alpha 
	\frac{ \braket{ii}{V}{\alpha\bar{\alpha}}  }{\epsilon_i+\epsilon_i-\epsilon_\alpha-\epsilon_{\bar{\alpha}}}
                  \sum_\gamma \braket{\alpha\bar{\alpha}}{V}{\gamma\bar{\gamma}} t_{ii}^{\gamma\bar{\gamma}}
\text.
\end{split}
\end{equation}

We stress that one important feature of the UEG model consists in the fact,
that enlarging the basis set does not alter the occupied and virtual orbitals.
We now examine the proposed approximations numerically. We choose the union of all
virtual states to be a very large number of 30046 states, which can be
considered a good approximation to the CBS limit for the present system.
In the following we gradually increase
the number of virtual states in the finite basis and evaluate the
approximate expressions for $\epsilon_{ii}^\textrm{B}$ and
$\epsilon_{ii}^\textrm{C}$ in Eqs.~(\ref{eqn:epsbapprox}) and
(\ref{eqn:epscapprox}) and compare them to the exact result in
Eq.~(\ref{eqn:uegabcd}).  The results for increasing numbers of virtual states are
given in Table ~\ref{tab:uegbcd}. The contribution of $\epsilon_{ii}^\textrm{B}$
is roughly twice as large as  $\epsilon_{ii}^\textrm{C}$.
We find that both energy contributions can be
approximated with remarkable accuracy using the presented expressions.
Although the approximations made for $\epsilon_{ii}^\textrm{B}$ and
$\epsilon_{ii}^\textrm{C}$ differ,
we can not observe any significant differences in
the accuracy of both terms.
The term $\epsilon_{ii}^\textrm{D}$, for which no
approximation was introduced, converges considerably faster, when compared
to the other two contributions $\epsilon_{ii}^\textrm{B}$ and
$\epsilon_{ii}^\textrm{C}$.
Hence, the BSIE of the PPL contribution can be reduced by a large portion successfully.
It appears that the remaining deviation is roughly in the same order of magnitude as the
rest contribution.
Therefore, it is a reasonable approximation to neglect both contributions from
$\epsilon_{ii}^\textrm{D}$ and $\Delta E^\textrm{rest}$.

For the above analysis, we have 
employed the fully converged CCSD amplitudes expanded in a basis of 30046 virtual states.
The amplitudes have been partitioned according to the cutoff $k_1$ into
sets corresponding to $t_{ii}^{\alpha\bar{\alpha}}$ and $t_{ii}^{a \bar{a}}$, which have been used
to compute $\epsilon_{ii}^\textrm{B}$ and $\epsilon_{ii}^\textrm{C}$.
However, in practice and for the following benchmark systems, we employ only CCSD amplitudes
that have been calculated using a finite virtual orbital basis set.

\section{Computational details}

In the following sections we present results obtained for a set of benchmark
systems including 107 molecules and atoms.  We employ aug-cc-pVXZ basis sets
for first-row elements and aug-cc-pV(X+d)Z basis sets for second-row
elements~\cite{Dunning2001, Kendall1992}.  These basis sets will be denoted as
AVXZ throughout this work. We obtained the reference energies using the quantum
chemistry package PSI4 \cite{Psi4ref}. We have modified the code such that the
$E^\textrm{ppl}$ contribution is extracted from the calculation as described in
Ref.~\cite{Irmler2019b}. For the CBS limit estimates we use AV5Z and AV6Z
energies and the extrapolation formula $E_X = E_\textrm{CBS} + a/X^3$, with the
basis set cardinal number $X$. This formula is used to get CBS estimates of all
three individual terms: $E^\textrm{mp2}$, $E^\textrm{ppl}$, and
$E^\textrm{rest}$. We use unrestricted Hartree--Fock orbital functions
and corresponding CCSD implementations for all open-shell systems.
All correlation energy calculations in this work used the frozen core approximation.

In addition, (F12*) calculations are performed using
TURBOMOLE~\cite{TURBOMOLE,Balasubramani2020,Bachorz2011,Haettig2010} and the
AVDZ, AVTZ, and AVQZ basis sets. We employ default settings, however, we use
the RI basis aug-cc-pV5Z developed by H{\"a}ttig~\cite{Haettig2005} in all
calculations.  We note that these large RI basis sets are employed for all types
of auxiliary functions in the TURBOMOLE implementation, {\it i.e.} \$cbas,
\$jkbas, and \$cabs. All results in the main text employ $\gamma=1.0$ in the
parametrization of the correlation factor. Further results using a different
$\gamma$ parameter can be found in the supplement information.

The derived approximate BSIE corrections to the PPL term were
obtained using our own coupled-cluster code cc4s, LIBINT2~\cite{Libint2} and
CTF~\cite{Solomonik2014}. In these calculations, the Hartree--Fock ground state
wave function was obtained with the NWChem package~\cite{Valiev2010} and interfaced to
cc4s as described in Ref.~\cite{Gallo2021}. We note that the introduced basis set correction
scheme requires consistent sets of occupied orbitals for varying basis sets.
However, this is not automatically guaranteed for degenerate sets of occupied orbitals.
In this work, we avoid arbitrary unitary rotations among degenerate sets of orbitals by
introducing point charges far away from the molecules and atoms that break
corresponding symmetries, lifting all possibly problematic degeneracies.
These point charges are sufficiently far away
to ensure that all computed correlation energies change by a numerically negligible small
amount.

\begin{figure}[t]
\begin{center}
\includegraphics[width=0.99\linewidth]{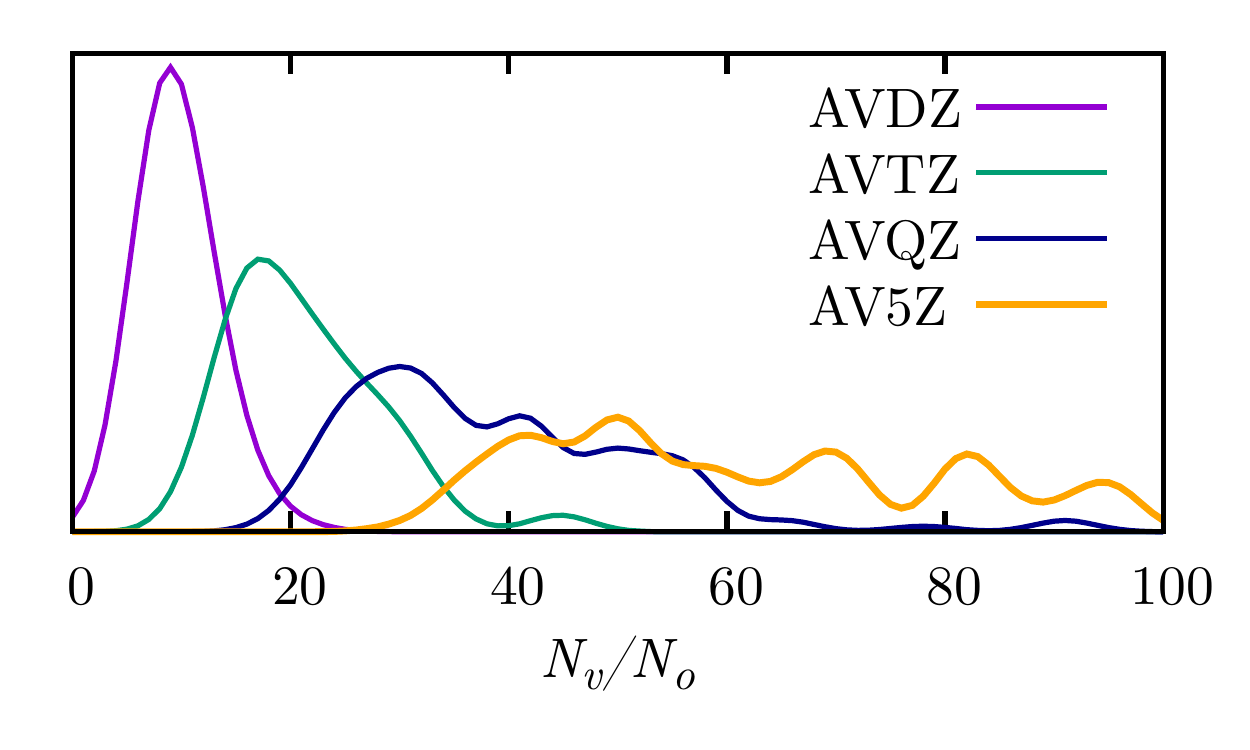}
\end{center}
\caption{\label{fig:nvnoratio}
Distribution of the number of virtual orbitals per occupied
orbital for all 107 studied systems when employing an atom-centered AVXZ basis set.
The same Gaussian function was used to smear the data
points.
  }
\end{figure}

For the newly introduced basis set correction scheme we construct frozen
natural orbitals (FNOs) on the level of second-order perturbation theory
\cite{Sosa1989,Klopper1997,Taube2005}. We truncate the virtual space used for
the CCSD calculations by choosing only $N_v$ FNOs with the largest occupation
number, where $N_v = X_\textrm{no} \times \text{max}\left(N_{o,\alpha},
N_{o,\beta}\right)$ with $X_\textrm{no} \in [12, 16, 20, 24, 28, 32]$.
$N_{o,\alpha}$ and $N_{o,\beta}$ refer to the number of occupied spin-up and
spin-down orbitals, respectively.  We stress that we use large basis sets (AVQZ
and AV5Z) for the construction of FNOs.
Therefore, the number of virtual orbitals, $N_v$, is defined differently
than for conventional quantum chemical calculations. In conventional
calculations with atom-centered basis sets, the total number of orbitals is
independent of the number of occupied orbitals but depends only on the atomic
species for a chosen basis set.  Yet, we seek to compare the BSIEs of
correlation energies calculated using both approaches. To provide an estimate
for which cardinal number in the AVXZ basis set family corresponds on average to
which number of virtual orbitals per occupied orbital, Fig.~\ref{fig:nvnoratio}
depicts $N_v/N_o$ for all studied atomic and molecular systems employing
conventional AVXZ (X=D,T,Q,5) basis sets.  We find that AVDZ and AVTZ roughly
correspond to $X_\textrm{no}=12$ and $X_\textrm{no}=20$, respectively.  Later,
it will be numerically verified that our choice of fixed number of virtuals per
occupied leads to a well-balanced energy description for different reactants.

We have calculated the correlation energies of in total 107 molecules and
atoms. Thereupon we evaluated 26 closed-shell reaction energies (REc), 39 open-shell
reaction energies (REo), 44 atomization energies (AE), 16 electron affinities (EA), and
22 ionization potentials (IP). This benchmark set is a subset of the one
studied by Knizia {\it et al.}~\cite{Knizia2009}. We had to exclude a number of
molecules from their benchmark set as some of the molecules have been too large
to be treated with our workflow. For some other molecules, we were not able to
converge to a common HF ground state with neither of the three packages NWChem, TURBOMOLE,
and PSI4. These molecules have also been excluded from our benchmark.  A
detailed list of the calculated molecules and corresponding reactions can be found
in the supplement information.

\section{Results}

This section presents results for molecular systems and is organized as follows.
In section~\ref{sec:total} we assess the convergence of the
diagrammatically decomposed CCSD correlation energy contributions,
confirming that the BSIEs in the total energy are dominated by the MP2 and PPL terms.
In section~\ref{sec:ediff} we show that this behavior persists
for most quantities computed from the total energies including reaction energies,
atomization energies, ionization potentials and electron attachment energies.
In addition, we explore the accuracy of the derived approximate correction to
the BSIE of the PPL term for all investigated quantities.
In section~\ref{sec:focal} we assess the accuracy of the corrected total CCSD
energies and related quantities using two practical settings for the introduced
focal-point approach and the respective BSIE corrections to the PPL term.
The obtained results are compared to conventional CCSD and CCSD(F12*) approaches.
Section~\ref{sec:triples} discusses our findings for the (T) and (T*) correlation
energy contributions using FNOs.

\subsection{Total energies}
\label{sec:total}
\begin{figure}[t]
\begin{center}
\includegraphics[width=0.99\linewidth]{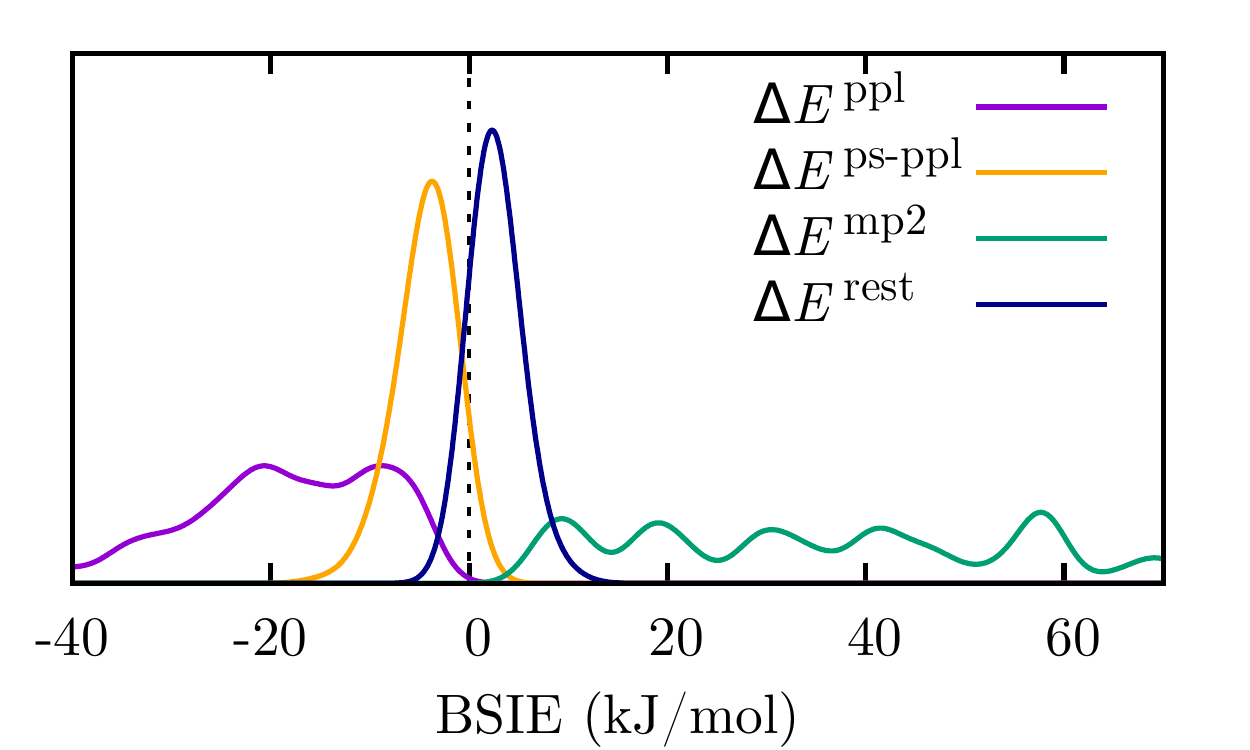}
\end{center}
\caption{\label{fig:channel}
Distribution of the basis set incompleteness error (BSIE)
of  various investigated energy channels (MP2, PPL and rest)
including the corrected PPL energy ($\Delta E^\textrm{ps-ppl}$) for 107 studied systems.
The energies were calculated using 16 frozen natural orbitals per occupied orbital
and are referenced to [56] values. The same Gaussian function was used to smear the data
points.
}
\end{figure}

We begin the analysis of the molecular systems by presenting results for the
basis set errors of the diagrammatically decomposed correlation energy
contributions for 107 molecules and atoms.  Fig.~\ref{fig:channel} depicts the
BSIEs of the PPL ($\Delta E^\textrm{ppl}$), MP2 ($\Delta E^\textrm{mp2}$) and
rest ($\Delta E^\textrm{rest}$) contributions.  Furthermore, the BSIEs of the
PPL energies corrected according to Eq.(\ref{eqn:Epsppl}) are also depicted
($\Delta E^\textrm{ps-ppl}$). The BSIEs are estimated using reference values
obtained from a [56] extrapolation.  The correlation energies are evaluated
using a Hartree--Fock reference wave function and 16 FNOs per occupied orbital
to approximate the virtual orbital manifold. This basis set size is on average
between AVDZ and AVTZ, as can be seen in Fig.~\ref{fig:nvnoratio}.  For the
construction of the FNOs, the one-particle reduced density matrix was
calculated in an AV5Z basis set.  Our findings show that MP2 energies
calculated using 16 FNOs per occupied orbital exhibit by far the largest BSIEs
when compared to the other contributions.  In contrast to MP2,
$E^\textrm{rest}$ is significantly better converged.  This analysis reveals
that $E^\textrm{rest}$ can already be well approximated using a smaller number
of natural orbitals than required for $E^\textrm{ppl}$ and $E^\textrm{mp2}$.
However, adding the basis set correction to $E^\textrm{ppl}$ as defined in
Eq.(\ref{eqn:Epsppl}), significantly reduces the remaining BSIE such that
$\Delta E^\textrm{ps-ppl}$ becomes comparable to $\Delta E^\textrm{rest}$ for
all studied systems. This demonstrates impressively that the approximation
derived in Sec.~\ref{sec:theory} can transfer its accuracy from the uniform
electron gas model system to real atoms and molecules.

\subsection{Energy differences}
\label{sec:ediff}

\begin{table*}[ht]
\caption{
BSIEs of correlation energy contributions to REc, REo and AEs.
Shown are the rms deviations to the [56] reference.
Results have been obtained using 12-32 FNOs per occupied orbitals
($X_\textrm{no}$), [23]-[45] extrapolations, and for
different conventional basis sets ranging from AVDZ up to AV6Z.
}
\label{tab:molchannelstat}
\begin{tabular*}{\textwidth}{@{\extracolsep{\fill}}lcccccccccccc}
\toprule
&
\multicolumn{4}{c}{REc (kJ/mol)}  &
\multicolumn{4}{c}{REo (kJ/mol)}    &
\multicolumn{4}{c}{AEs (kJ/mol)}
\\
       &
\dmp   &
\dppl  &
\drest &
\dpsppl &
\dmp   &
\dppl  &
\drest &
\dpsppl &
\dmp   &
\dppl  &
\drest &
\dpsppl
\\
\midrule
$X_\textrm{no}$=12   &  6.396 & 1.850 &  1.212 & 1.199 &  28.325 &  8.811 &  1.513 & 1.986 &  39.350 & 14.985 &  2.783 & 2.257 \\
$X_\textrm{no}$=16   &  4.576 & 1.302 &  0.822 & 0.931 &  20.556 &  6.509 &  0.783 & 1.502 &  28.230 & 11.118 &  1.437 & 1.321 \\
$X_\textrm{no}$=20   &  2.978 & 0.877 &  0.437 & 0.739 &  15.771 &  5.224 &  0.489 & 1.390 &  22.094 &  8.941 &  1.005 & 0.949 \\
$X_\textrm{no}$=24   &  2.240 & 0.540 &  0.338 & 0.667 &  12.429 &  4.130 &  0.308 & 1.257 &  17.231 &  7.115 &  0.650 & 0.723 \\
$X_\textrm{no}$=28   &  2.289 & 0.626 &  0.299 & 0.536 &   9.416 &  3.114 &  0.242 & 1.284 &  14.024 &  5.930 &  0.600 & 0.664 \\
$X_\textrm{no}$=32   &  2.476 & 0.649 &  0.516 & 0.485 &   7.241 &  2.398 &  0.392 & 1.325 &  11.494 &  4.903 &  0.559 & 0.627 \\ 
\midrule
AVDZ     & 15.207 & 2.400 &  3.468 &   -   &  43.096 & 10.296 &  5.200 &   -   &  67.192 & 21.166 & 12.269 &   -   \\
AVTZ     &  6.981 & 1.767 &  2.040 &   -   &  20.598 &  5.675 &  2.544 &   -   &  27.203 &  9.567 &  4.153 &   -   \\
AVQZ     &  3.212 & 1.077 &  0.653 &   -   &   8.945 &  2.666 &  0.624 &   -   &  11.866 &  4.321 &  0.831 &   -   \\
AV5Z     &  1.885 & 0.640 &  0.288 &   -   &   4.553 &  1.370 &  0.300 &   -   &   6.122 &  2.257 &  0.274 &   -   \\
AV6Z     &  1.092 & 0.371 &  0.164 &   -   &   2.636 &  0.793 &  0.177 &   -   &   3.541 &  1.306 &  0.158 &   -   \\
\midrule
%
$[$23$]$ &  5.420 & 1.708 &  2.049 &   -   &  11.576 &  3.901 &  2.852 &   -   &  11.131 &  4.941 &  2.534 &   -   \\
$[$34$]$ &  2.007 & 0.842 &  0.815 &   -   &   1.393 &  0.662 &  1.468 &   -   &   1.237 &  0.708 &  1.845 &   -   \\
$[$45$]$ &  0.730 & 0.262 &  0.148 &   -   &   0.719 &  0.237 &  0.400 &   -   &   0.458 &  0.172 &  0.416 &   -   \\
\bottomrule
\end{tabular*}

\end{table*}

More decisive than well converged total energies is the question of how the
proposed correction scheme works for energy differences.  Therefore, we analyze
the BSIEs for the different channels ($E^\textrm{ppl}$, $E^\textrm{mp2}$, and
$E^\textrm{rest}$) for REc, REo, and AEs. The results are summarized in
Table~\ref{tab:molchannelstat} for increasing numbers of FNOs as well as for
the basis sets AVDZ-AV6Z. The MP2 contribution shows the largest BSIE followed
by the PPL contribution. This is in accordance with the findings for the total
energies, discussed in the previous section. We stress that only in the case of
REc, the BSIE of $E^\textrm{rest}$ and $E^\textrm{ppl}$ is of comparable
magnitude.  Furthermore, we note that the computed errors using FNOs for some
systems  become larger again or do not reduce significantly for
$X_\textrm{no}$\textgreater24. We attribute this behavior to not sufficiently
well converged FNOs. When approaching $X_\textrm{no}$\textgreater24, one would
require even bigger basis sets than the employed AV5Z for the construction of
the FNOs. Generally, it is not expected that the errors are significantly
smaller than when using all possible virtual orbitals in the AV5Z basis set.

Especially for the large basis sets, the BSIE of the rest contribution is
remarkably small; the rms deviation for AV5Z is only around 0.3~kJ/mol and lower.
A similar high accuracy can be attained when using only a comparably small number
of 20 FNOs per occupied orbital, which achieves rms deviations of around 0.5~kJ/mol
for the reaction energies and 1~kJ/mol for atomization energies.

For REo and AEs the PPL contribution converges significantly slower with respect to the
basis set size compared to  $E^\textrm{rest}$. Furthermore, the BSIE cannot be diminished
considerably with a finite number of FNOs.  This behavior changes when taking
the proposed correction into account. Compared to the uncorrected PPL contribution,
the BSIE of the corrected PPL contribution is
reduced by a factor of four and more, when using only 20 FNOs per occupied
or less. For REc the correction has no significant effect.

In summary, rms deviations of the rest contributions (\drest) and corrected PPL (\dpsppl)
contributions are on the scale of 1~kJ/mol when using 20 FNOs per occupied
orbital. Reaching a similar accuracy by employing conventional basis set calculations
would require a [34] extrapolation.

\subsection{Benchmarking a practical focal-point approach}
\label{sec:focal}

Based on the findings in the previous sections we now define and assess a practical
focal-point approach to compute total CCSD energies. For an even-tempered composition
we combine extrapolated MP2 energies with CCSD calculations employing FNOs. We
introduce the following two compositions:
\begin{equation}
\begin{split}
 E^\textrm{FPa} = & E^\textrm{mp2} ([34]) + E^\textrm{rest} (12) + E^\textrm{ppl}(12)  \\
	        = & E^\textrm{ccsd} (12) - E^\textrm{mp2} (12) + E^\textrm{mp2} ([34])
\end{split}
\end{equation}
and
\begin{equation}
\begin{split}
	E^\textrm{FPb} = & E^\textrm{mp2} ([45]) + E^\textrm{rest} (20) + E^\textrm{ppl}(20) \\
	        = & E^\textrm{ccsd} (20) -  E^\textrm{mp2} (20) + E^\textrm{mp2} ([45]) \text. 
\end{split}
\end{equation}
$E^\textrm{mp2} (X_\textrm{no})$, $E^\textrm{ccsd} (X_\textrm{no})$,
$E^\textrm{rest} (X_\textrm{no})$ and  $E^\textrm{ppl}(X_\textrm{no})$
refer to the corresponding correlation energy
contributions calculated employing $X_\textrm{no}$ FNOs per occupied orbital.
$E^\textrm{mp2} ([34])$ and  $E^\textrm{mp2} ([45])$ refer to MP2 correlation
energies obtained from a [34] and [45] extrapolation, respectively.
For the first ansatz ($E^\textrm{FPa}$) we construct the FNOs using an
AVQZ calculation, whereas for the second ansatz ($E^\textrm{FPb}$) the AV5Z basis set is
used. In this section we will explore benchmark results obtained using both approaches
with and without the introduced \sdppl~ correction that depends on the respective
pair-specific extrapolated MP2 energies and correlation hole depths.
The corresponding BSIEs are summarized in Table~\ref{tab:focalstat}.

\begin{table*}[t]
\caption{
CCSD valence correlation energy basis set incompleteness error for closed-shell reaction (REc), open-shell
reactions (REo), atomization energies (AEs), ionization potentials (IPs), and
electron affinities (EAs).
Reference is obtained from a [56] extrapolation. Two different variants
of the focal-point approximation are used with and without correction.
Details are found in the main text.
}
\label{tab:focalstat}
\begin{tabular*}{\textwidth}{@{\extracolsep{\fill}}lcccccccccc}
\toprule
&
\multicolumn{2}{c}{REc (kJ/mol)} &
\multicolumn{2}{c}{REo (kJ/mol)} &
\multicolumn{2}{c}{AEs (kJ/mol)} &
\multicolumn{2}{c}{IPs (kJ/mol)} &
\multicolumn{2}{c}{EAs (kJ/mol)}
\\
    &
max &
rms &
max &
rms &
max &
rms &
max &
rms &
max &
rms \\
\midrule
%
FPa               &   4.823 &  1.905 &   15.403 &   7.130  &   24.223 & 11.786  & 10.244 &  3.301 & 15.384 &  5.414 \\
FPb               &   2.828 &  1.029 &   10.998 &   5.445  &   16.660 &  8.461  &  6.837 &  2.652 &  9.203 &  3.766 \\
FPa+ \sdppl &   7.248 &  2.541 &    8.836 &   3.024  &    9.017 &  2.400  &  3.359 &  1.520 &  5.666 &  1.603 \\ 
FPb+ \sdppl &   3.253 &  1.102 &    2.424 &   1.049  &    3.491 &  0.848  &  1.475 &  0.733 &  3.062 &  0.828 \\
\midrule
$[$23$]$          &  15.637 &  5.814 &   33.638 &  10.460  &   26.000 &  8.443  &  8.814 &  4.366 &  6.170 &  3.033 \\ 
$[$34$]$          &   6.222 &  2.031 &    4.545 &   1.938  &    3.965 &  1.874  &  1.993 &  0.828 &  1.772 &  0.895 \\
$[$45$]$          &   1.137 &  0.438 &    1.864 &   0.650  &    1.302 &  0.532  &  0.213 &  0.112 &  0.181 &  0.113 \\
\midrule
%
(F12*) @ AVDZ     &  9.859 &  3.163 &   14.625 &   4.274  &   11.957 &  5.821  &  8.375 &  4.632 &  7.059 &  4.720 \\
(F12*) @ AVTZ     &  3.838 &  1.434 &    5.946 &   1.657  &    3.834 &  1.520  &  2.685 &  1.393 &  1.810 &  1.189 \\
(F12*) @ AVQZ     &  2.233 &  0.869 &    1.727 &   0.719  &    1.795 &  0.532  &  0.898 &  0.543 &  1.015 &  0.472 \\
%
%
\bottomrule
\end{tabular*}

\end{table*}

The uncorrected focal-point approaches $E^\textrm{FPa}$ and $E^\textrm{FPb}$
yield only satisfying BSIEs for the closed-shell reaction energies. Here, FPa achieves the
quality of the [34] result, although the CCSD calculation is performed with a
significantly smaller virtual space of only 12 FNOs per occupied orbital. For the 
open-shell reactions and other properties, the focal-point method performs
significantly worse with rms deviations between 3-12~kJ/mol and a maximum error of around 25~kJ/mol.

The focal-point approaches including the \sdppl~correction yield significantly more consistent
BSIEs for all studied energy differences. The rms deviations are 1.5-3~kJ/mol and around 1~kJ/mol
for FPa and FPb, respectively. For the FPb+\sdppl~approach the maximum deviation is
below 4~kJ/mol for all considered reactions.

We note that for the closed-shell reactions the corrected focal-point results
show larger rms deviations and larger maximum errors than the uncorrected
variants.  We attribute this to fortuitous error cancellation between the
individual energy contributions to the CCSD correlation energy. This is only
visible when the \sdppl-correction is insignificant as it is the case for the
closed-shell reaction energies (see Sec.~\ref{sec:ediff}).  Correcting for the
BSIE in the PPL term, reduces this error compensation, causing slightly larger
BSIEs for closed-shell reaction energies. However, the results for REc
obtained including the {\sdppl} correction are of comparable size to open-shell
reactions and other properties.

The extrapolation using the AVDZ and AVTZ basis sets shows large maximum errors of
up to 30~kJ/mol and the rms deviation ranges from 3 to 10~kJ/mol. The [34]
extrapolation yields satisfying results with rms deviations of  around
2~kJ/mol, for IPs and EAs already below 1~kJ/mol.  Although the
[45] extrapolation yields the best statistical results, a CCSD calculation
with the large AV5Z basis set is only possible for small systems.  We stress
that more sophisticated extrapolation techniques were already  tested for the
original version of the employed benchmark set. Results can be found in the
supplement of Ref.~\cite{Knizia2009}. Knizia \textit{et al.} conclude that ``[\ldots] in our
benchmark set there are only few systems where using either extrapolation
scheme makes a noteworthy difference``.  Thus, the corrected FPa ansatz is to
be preferred over [23] extrapolation and the corrected FPb seems to be superior
compared with the [34] extrapolation. We stress that in both cases larger HF
and MP2 calculations have to be performed in order to obtain the final result.

For comparison, CCSD(F12*) results are also given for three different basis sets.
The F12 results obtained using the AVQZ basis set reach almost the quality of the
[45] extrapolation, with rms deviations well below 1~kJ/mol.
For the (F12*)@AVTZ results, the rms deviations are only around
1.5~kJ/mol, whereas (F12*)@AVDZ yields results that show rms deviations with about 3-5~kJ/mol.
We note that (F12*)@AVTZ yields results with smaller rms deviations than the [34] extrapolation.
%
%
%
We note that the size of the virtual space in the CCSD calculation for
(F12*)@AVDZ and FPa+\sdppl~is similar. The same is true for (F12*)@AVTZ and
FPb+\sdppl. However, the FPa and FPb approaches require HF and MP2
calculations using the AVQZ and AV5Z basis sets, respectively.  Therefore, the
entire computational cost of the proposed focal-point approaches depends
strongly on the efficiency of the employed HF and MP2 algorithms.
Further statistical analysis of the test set using HF and conventional CCSD 
is provided in the supplementary information.

\subsection{Perturbative triples contribution}
\label{sec:triples}

\begin{table*}[t]
\caption{
BSIE of the (T) contribution to closed-shell reaction (REc), open-shell
reactions (REo), atomization energies (AEs), ionization potentials (IPs), and
electron affinities (EAs). Shown are the rms deviations to the [56] reference.
}
\label{tab:triples}
\begin{tabular*}{\textwidth}{@{\extracolsep{\fill}}lcccccccccc}
\toprule
                                 &
\multicolumn{2}{c}{REc (kJ/mol)} &
\multicolumn{2}{c}{REo (kJ/mol)} &
\multicolumn{2}{c}{AEs (kJ/mol)} &
\multicolumn{2}{c}{IPs (kJ/mol)} &
\multicolumn{2}{c}{EAs (kJ/mol)}
\\
     &
(T)  &
(T*) &
(T)  &
(T*) &
(T)  &
(T*) &
(T)  &
(T*) &
(T)  &
(T*)
\\
\midrule
$X_\textrm{no}$=12 & 0.914 & 0.522 & 1.925 & 0.456 & 2.881 & 0.633 & 0.808 & 0.387 & 1.378 & 0.710 \\
$X_\textrm{no}$=16 & 0.583 & 0.385 & 1.175 & 0.316 & 1.765 & 0.350 & 0.522 & 0.252 & 0.877 & 0.415 \\
$X_\textrm{no}$=20 & 0.418 & 0.265 & 0.808 & 0.255 & 1.336 & 0.259 & 0.381 & 0.177 & 0.679 & 0.333 \\
$X_\textrm{no}$=24 & 0.292 & 0.203 & 0.616 & 0.227 & 0.989 & 0.236 & 0.298 & 0.122 & 0.506 & 0.207 \\
$X_\textrm{no}$=28 & 0.268 & 0.183 & 0.487 & 0.241 & 0.824 & 0.209 & 0.260 & 0.110 & 0.421 & 0.166 \\
$X_\textrm{no}$=32 & 0.206 & 0.233 & 0.397 & 0.244 & 0.699 & 0.192 & 0.223 & 0.092 & 0.364 & 0.129 \\
\midrule
AVDZ               & 1.976 & 3.231 & 5.176 & 3.216 & 8.176 & 2.826 & 3.139 & 2.364 & 4.330 & 2.676 \\
AVTZ               & 1.055 & 0.823 & 1.485 & 0.997 & 2.165 & 0.450 & 0.862 & 0.432 & 1.275 & 0.566 \\
AVQZ               & 0.509 & 0.269 & 0.667 & 0.433 & 0.913 & 0.153 & 0.366 & 0.185 & 0.571 & 0.277 \\
AV5Z               & 0.270 & 0.135 & 0.303 & 0.231 & 0.425 & 0.087 & 0.179 & 0.088 & 0.296 & 0.151 \\
AV6Z               & 0.157 & 0.078 & 0.191 & 0.134 & 0.246 & 0.049 & 0.103 & 0.050 & 0.171 & 0.087 \\
\midrule
$[$23$]$           & 0.888 &   -   & 0.705 &   -   & 0.570 &   -   & 0.232 &   -   & 0.170 &   -   \\
$[$34$]$           & 0.255 &   -   & 0.188 &   -   & 0.133 &   -   & 0.066 &   -   & 0.121 &   -   \\
$[$45$]$           & 0.035 &   -   & 0.088 &   -   & 0.093 &   -   & 0.024 &   -   & 0.040 &   -   \\
\bottomrule
\end{tabular*}

\end{table*}

Having assessed the introduced focal-point approach for the CCSD method,
we now turn to the discussion of BSIEs in the perturbative triples contribution
to the CCSD(T) energies calculated using FNOs.
In addition to the conventional approach of computing the (T) contribution,
we will also explore the (T*) approximation, which approximates
the CBS limit of (T) by rescaling the finite basis set result with a factor
estimated on the level of MP2 theory as outlined in Ref.~\cite{Knizia2009}.
In this work, the scaling factor corresponds to  $E^\textrm{mp2} ([45])/E^\textrm{mp2} (X_\textrm{no})$.
The results are summarized in Table~\ref{tab:triples}. The rms deviations for the AVDZ are
2-8~kJ/mol and even with the corrected values, denoted as (T*), the errors are
within the range of 2-3~kJ/mol.
With increasing cardinal numbers the deviations reduce considerably.
AVQZ results show deviations of up to 1~kJ/mol,
reducing even further for the (T*) correction where they
do not surpass 0.5~kJ/mol.
As it is apparent from the presented data, the usage of FNOs together
with the (T*) ansatz seems to be highly effective.
Already 12 FNOs per occupied orbital suffice to reduce the rms deviations
to 0.7~kJ/mol and lower.
When using 20 FNOs instead, this deviation reduces smoothly below
0.35~kJ/mol.
We stress that computing the (T*) scaling factor using a [56] extrapolation instead
of [45] extrapolation has almost no effect (0.05~kJ/mol in the rms BSIEs).

Considering the findings for the BSIEs listed in Tabs.~\ref{tab:focalstat} and \ref{tab:triples}
in combination, shows that it is possible to obtain CCSD(T) correlation energy estimates of
REc, REo, AEs, IPs and EAs with a root mean square deviation from the CBS limit below 4~kJ/mol
using 12 FNOs per occupied orbital only.
Employing 20 FNOs per occupied orbital reduces the rms BSIE to around 1~kJ/mol
for all computed energy differences.
A detailed summary of all computed energies and BSIEs can be found in the
supplement information.


\section{Summary and conclusion}

In this work we introduced a new CCSD basis set correction scheme that employs FNOs and
exhibits an excellent trade-off between virtual orbital basis set size and remaining BSIE.
The introduced correction aims at removing the BSIEs of the most important
contributions to the CCSD correlation energy in the CBS limit originating from the MP2 and PPL terms.
We stress that our approach is electron pair-specific. 
The presented results for reaction energies, atomization energies, ionization potentials
and electron attachment energies exhibit a rapid convergence with respect to the basis set size.
Furthermore, we have shown that the CBS limit of the (T) contribution to the CCSD(T) energy can be
approximated with a similar efficiency when employing FNOs and a rescaling procedure previously
referred to as (T*)~\cite{Knizia2009}.

The introduced focal-point approach is an interesting alternative to conventional basis set truncation
techniques that are based on cardinal numbers. FNOs give access to much more finely
incremented basis set sizes, while maintaining a stable and rapid convergence of many
properties to the CBS limit. We did not observe significant shell filling effects
that might cause non-monotonic energy convergence with respect to the FNO basis set size.
However, we note that the presented approach relies on the availability of computational efficient
algorithms to compute MP2 energies and FNOs.

We stress that our approach is directly transferable to
{\em ab initio} calculations employing pseudo potentials.
This will be beneficial for solid state calculations in a plane wave basis set,
which is one of the main motivations for this research.
Most solid state calculations using coupled-cluster theory and
plane-wave basis sets performed so far
employ FNOs and the electronic transition structure factor needed
to compute correlation hole depths is readily available~\cite{Kresse2011,Liao2016,Gruber2018}. 
Furthermore, we stress that the outlined basis set corrections will also be interesting
to other many-electron theories, where similar interference effects
play an important role, for instance, Auxiliary-Field Quantum Monte Carlo~\cite{Morales2020}.

Finally, we note that the introduced basis set correction scheme is expected to work reliably
in systems exhibiting correlation hole depths that vary strongly between different electron
pairs. This includes situations where core-valence electron correlation effects play an important role.

\section{Acknowledgements}
The authors thankfully acknowledge support and funding from the European
Research Council (ERC) under the European Unions
Horizon 2020 research and innovation program (Grant Agreement No 715594).
The computational results presented have been achieved in part using
the Vienna Scientific Cluster (VSC).

\bibliography{article.bib}

\onecolumngrid
\newpage
\part*{Supplementary information}
\appendix

\section{Introduction}

This document presents additional data for the manuscript.  Energies of the
individual atoms and molecules, as well as the output of the psi4 calculations
can be found in the online supplementary information in \cite{supp}.

Since the geometries for all considered systems can be found in the supplementary
material of Knizia {\it et al.}~\cite{Knizia2009}, we do not reiterate them here.

\newpage
\section{F12 results}

\begin{table*}[ht]
\begin{center}
  \caption{%
    BSIEs of the valence correlation energy using the (F12*) method. Calculated results
    use the AV\{D,T,Q\}Z basis sets and \( \gamma \in \{1.0, 1.4\} \).
  }
\label{tab:si:molchannelstat}
\newcommand{\ra}[1]{\renewcommand{\arraystretch}{#1}}
\ra{1.2}
\begin{tabular}{@{}lccccccccccccccc@{}}
\toprule
&
& \multicolumn{2}{c}{REc (kJ/mol)}
&
&  \multicolumn{2}{c}{REo (kJ/mol)}
&
& \multicolumn{2}{c}{AEs (kJ/mol)}
&
& \multicolumn{2}{c}{IPs (kJ/mol)}
&
& \multicolumn{2}{c}{EAs (kJ/mol)} \\
                         &&   max   &  rms   &&   max    &  rms     &&  max     &  rms    &&   max  &   rms  &&  max   &    rms \\
\cmidrule{3-4}  \cmidrule{6-7} \cmidrule{9-10} \cmidrule{12-13} \cmidrule{15-16}
AVDZ, $\gamma=1.0$ &&   9.859 &  3.163 &&   14.625 &   4.274  &&   11.957 &  5.821  &&  8.375 &  4.632 &&  7.059 &  4.720 \\
AVDZ, $\gamma=1.4$ &&  13.834 &  3.885 &&   10.181 &   3.870  &&   17.937 &  6.829  && 14.019 &  7.040 && 12.439 &  7.425 \\
\midrule
AVTZ, $\gamma=1.0$ &&   3.838 &  1.434 &&    5.946 &   1.657  &&    3.834 &  1.520  &&  2.685 &  1.393 &&  1.810 &  1.189 \\
AVTZ, $\gamma=1.4$ &&   2.931 &  1.004 &&    5.142 &   1.512  &&    6.138 &  1.762  &&  3.250 &  1.582 &&  2.845 &  1.641 \\
\midrule
AVQZ, $\gamma=1.0$ &&   2.233 &  0.869 &&    1.727 &   0.719  &&    1.795 &  0.532  &&  0.898 &  0.543 &&  1.015 &  0.472 \\
AVQZ, $\gamma=1.4$ &&   1.823 &  0.708 &&    1.217 &   0.555  &&    1.326 &  0.383  &&  0.533 &  0.299 &&  0.885 &  0.353 \\
\bottomrule
\end{tabular}

\end{center}
\end{table*}

\newpage
\section{Total energies for chemical reactions}

\begin{table}[H]
  \centering
  \caption{%
    Hartree--Fock (HF) energy contribution obtained with AV6Z, valence correlation
    energy from CCSD, and (T) contribution, both obtained from [56]-extrapolation,
    of the reaction energies for closed-shell systems. All energies are given in kJ/mol.
  }
  \label{tab:si:reactionclosed}
  \begin{tabular}{lrrr}
  \toprule
  Reaction                                   &     HF   &    CCSD    & (T)  \\
  \midrule
  CO + H$_2$ $\rightarrow$  HCHO                        &     1.058 &    -23.427 &      0.727 \\
  CO + H$_2$O $\rightarrow$  CO$_2$ + H$_2$             &     0.242 &    -16.963 &    -10.158 \\
  CH$_3$OH + HCl $\rightarrow$  CH$_3$Cl + H$_2$O       &   -25.113 &     -6.664 &     -1.830 \\
  H$_2$O + CO $\rightarrow$  HCOOH                      &    -7.588 &    -26.095 &     -4.056 \\
  CH$_3$OH + H$_2$S $\rightarrow$ CH$_3$SH + H$_2$O     &   -32.176 &    -11.292 &     -2.132 \\
  CS$_2$ + 2 H$_2$O $\rightarrow$ CO$_2$ + 2 H$_2$S     &  -122.291 &     57.265 &     17.649 \\
  HNCO + H$_2$O $\rightarrow$ CO$_2$ + NH$_3$           &   -96.839 &      9.601 &      1.302 \\
  CH$_4$ + Cl$_2$ $\rightarrow$ CH$_3$Cl + HCl          &  -110.008 &      8.814 &      3.027 \\
  Cl$_2$ + F$_2$ $\rightarrow$  2 ClF                   &  -142.439 &     22.963 &      6.220 \\
  CO + Cl$_2$ $\rightarrow$  COCl$_2$                   &   -56.532 &    -48.165 &    -10.090 \\
  CO$_2$ + 3 H$_2$ $\rightarrow$ CH$_3$OH + H$_2$O      &  -118.688 &    -14.650 &     15.493 \\
  HCHO + H$_2$ $\rightarrow$  CH$_3$OH                  &  -119.505 &     -8.186 &      4.608 \\
  CO + 2 H$_2$ $\rightarrow$  CH$_3$OH                  &  -118.447 &    -31.614 &      5.335 \\
  SO$_3$ + CO $\rightarrow$ SO$_2$ + CO$_2$             &  -159.381 &    -15.010 &     -6.383 \\
  H$_2$ + Cl$_2$ $\rightarrow$  2 HCl                   &  -213.351 &     15.976 &      6.784 \\
  C$_2$H$_2$ + H$_2$ $\rightarrow$ C$_2$H$_4$           &  -216.388 &      5.679 &      4.369 \\
  SO$_2$ + H$_2$O$_2$ $\rightarrow$ SO$_3$ + H$_2$O     &  -231.750 &     16.268 &      4.335 \\
  CO + 3 H$_2$ $\rightarrow$ CH$_4$ + H$_2$O            &  -246.902 &    -31.115 &      7.262 \\
  HCN + 3 H$_2$ $\rightarrow$ CH$_4$ + NH$_3$           &  -335.389 &      3.306 &     11.059 \\
  H$_2$O$_2$ +  H$_2$ $\rightarrow$ 2 H$_2$O            &  -391.373 &     18.221 &      8.110 \\
  CO + H$_2$O$_2$ $\rightarrow$ CO$_2$ + H$_2$O         &  -391.131 &      1.258 &     -2.048 \\
  2 NH$_3$ + 3 Cl$_2$  $\rightarrow$ N$_2$ + 6 HCl      &  -482.764 &     63.682 &     11.655 \\
  3 N$_2$H$_4$  $\rightarrow$ 4 NH$_3$ + N$_2$          &  -470.681 &     32.160 &     -0.570 \\
  H$_2$ + F$_2$ $\rightarrow$ 2 HF                      &  -610.736 &     32.890 &     12.878 \\
  CH$_4$ + 4 H$_2$O$_2$ $\rightarrow$ CO$_2$ + 6 H$_2$O & -1318.348 &     87.035 &     15.020 \\
  2 NH$_3$ + 3 F$_2$  $\rightarrow$ N$_2$ + 6 HF        & -1674.919 &    114.422 &     29.936 \\
  \bottomrule
\end{tabular}

\end{table}

\begin{table}[H]
  \centering
  \caption{%
    Open-shell systems (REo).
    The interpretation of the columns is as in Table~\ref{tab:si:reactionclosed}.
  }
  \begin{tabular}{lrrr}
  \toprule
  Reaction                                     &    HF     &    CCSD    & (T)  \\
  \midrule
  HCl + H $\rightarrow$ Cl + H$_2$                        &   -28.206 &     12.237 &      6.653 \\
  H$_2$O + F$_2$ $\rightarrow$  2 HF + O                  &  -310.135 &    232.071 &     27.904 \\
  OH + H$_2$ $\rightarrow$  H$_2$O + H                    &   -14.235 &    -45.143 &     -7.958 \\
  CO + OH $\rightarrow$  CO$_2$ + H                       &   -13.994 &    -62.106 &    -18.116 \\
  S+ 2 HCl $\rightarrow$ H$_2$S + Cl$_2$                  &    21.046 &   -123.055 &    -16.659 \\
  N + O$_2$ $\rightarrow$  NO + O                         &  -110.043 &    -30.128 &      0.832 \\
  4 HCl + O$_2$ $\rightarrow$ 2 H$_2$O + Cl$_2$           &   -63.503 &    -86.576 &     -4.193 \\
  2 NO $\rightarrow$  N$_2$ + O$_2$                       &  -141.430 &    -30.529 &     -3.550 \\
  2 H$_2$O$_2$ $\rightarrow$ 2 H$_2$O + O$_2$             &  -292.541 &     91.065 &      6.845 \\
  Cl$_2$ + H $\rightarrow$  HCl + Cl                      &  -241.557 &     28.214 &     13.437 \\
  2 SO$_2$ + O$_2$ $\rightarrow$  2 SO$_3$                &  -170.959 &    -58.530 &      1.825 \\
  Cl + OH $\rightarrow$  HOCl                             &   -43.344 &   -180.661 &    -21.899 \\
  H$_2$S + F$_2$ $\rightarrow$  S + 2 HF                  &  -418.431 &    139.968 &     22.753 \\
  2 NH$_2$ $\rightarrow$ N$_2$H$_4$                       &  -136.148 &   -153.938 &    -14.755 \\
  NO + N $\rightarrow$  O + N$_2$                         &  -251.473 &    -60.657 &     -2.717 \\
  O + 2 HCl $\rightarrow$ H$_2$O + Cl$_2$                 &   -87.251 &   -215.157 &    -21.810 \\
  SO$_2$ + O $\rightarrow$ SO$_3$                         &  -140.979 &   -201.134 &    -18.801 \\
  CS + O $\rightarrow$ CO + S                             &  -319.067 &    -54.285 &      6.553 \\
  CH$_3$OH + O $\rightarrow$ HCHO + H$_2$O                &  -181.096 &   -190.995 &    -19.633 \\
  NH + H $\rightarrow$  NH$_2$                            &  -277.962 &   -131.485 &     -5.679 \\
  Si + 2 H$_2$  $\rightarrow$  SiH$_4$                    &  -373.824 &    -64.824 &     -3.810 \\
  CS + S $\rightarrow$  CS$_2$                            &  -280.543 &   -143.489 &    -25.977 \\
  NH$_2$ + H $\rightarrow$  NH$_3$                        &  -348.191 &   -127.825 &     -6.023 \\
  2 H$_2$ + O$_2$ $\rightarrow$  2H$_2$O                  &  -490.204 &    -54.624 &      9.376 \\
  CO$_2$ + C $\rightarrow$  2 CO                          &  -425.583 &   -104.610 &     -9.449 \\
  C + H$_2$O $\rightarrow$  CO + H$_2$                    &  -425.342 &   -121.574 &    -19.607 \\
  N$_2$H$_4$ + O$_2$ $\rightarrow$ N$_2$ +2 H2O           &  -542.239 &    -33.402 &      3.387 \\
  2 NH + NH $\rightarrow$ N$_2$ + H$_2$                   &  -393.195 &   -288.506 &    -32.102 \\
  C + S$_2$ $\rightarrow$  CS$_2$                           &  -472.107 &   -224.499 &    -36.431 \\
  2 CO + 2 NO $\rightarrow$  N$_2$ + 2 CO$_2$             &  -631.152 &   -119.080 &    -14.490 \\
  CH$_4$ + 2 O$_2$ $\rightarrow$ CO$_2$ + 2 H$_2$O        &  -733.265 &    -95.096 &      1.331 \\
  4 NH$_3$ + 5 O$_2$ $\rightarrow$ 4 NO + 6 H$_2$O        &  -873.176 &    -71.306 &     17.830 \\
  2 NH$_3$ + 2 NO + O $\rightarrow$ 2 N$_2$ + 3 H$_2$O    &  -774.948 &   -268.581 &    -17.898 \\
  C + O$_2$ $\rightarrow$  CO$_2$                         &  -915.304 &   -193.161 &    -20.390 \\
  CS$_2$ + 3 O$_2$ $\rightarrow$ CO$_2$ + 2 SO$_2$        &  -900.329 &   -224.948 &    -13.949 \\
  CH$_4$ + 4 NO $\rightarrow$ 2 N$_2$ + CO$_2$ + 2 H$_2$O & -1016.126 &   -156.154 &     -5.768 \\
  CH$_4$ + NH$_3$ + 3 O $\rightarrow$ HCN + 3 H$_2$O      &  -566.415 &   -600.848 &    -56.136 \\
  2 C + H$_2$ $\rightarrow$  C$_2$H$_2$                   &  -875.468 &   -315.782 &    -35.796 \\
  4 NH$_3$ + 3 O$_2$  $\rightarrow$  2 N$_2$ + 6 H$_2$O   & -1156.037 &   -132.365 &     10.730 \\
  \bottomrule
\end{tabular}

\end{table}

\begin{table}[H]
  \centering
  \caption{%
    Atomization energies (AEs).
    The interpretation of the columns is as in Table~\ref{tab:si:reactionclosed}.
  }
  \begin{tabular}{lrrr}
  \toprule
  Reaction   &    HF    &    CCSD    & (T)  \\
  \midrule
  F$_2$                 & -151.770 &    281.165 &     31.299 \\
  Cl$_2$                &   81.149 &    147.632 &     20.090 \\
  ClF                   &   35.909 &    202.917 &     22.585 \\
  ClO                   &   25.919 &    217.176 &     26.213 \\
  NH                    &  215.115 &    126.599 &      4.605 \\
  CH                    &  238.401 &    108.830 &      3.647 \\
  S2                    &  215.312 &    185.460 &     30.731 \\
  OH                    &  286.366 &    154.038 &      7.068 \\
  HCl                   &  322.705 &    119.418 &      6.653 \\
  P$_2$                 &  155.695 &    286.663 &     41.507 \\
  O$_2$                 &  110.998 &    343.738 &     39.427 \\
  SO                    &  226.121 &    263.028 &     33.588 \\
  SiH$_2$(${}^{3}$B$_1$)&  446.960 &    108.617 &      1.951 \\
  HF                    &  404.939 &    177.728 &      9.210 \\
  NO                    &  221.041 &    373.866 &     38.595 \\
  SiH$_2$(${}^{1}$A$_1$)&  463.613 &    176.108 &      3.550 \\
  PH$_2$                &  456.840 &    182.551 &      4.918 \\
  HOCl                  &  329.710 &    334.699 &     28.967 \\
  CS                    &  406.876 &    266.470 &     41.186 \\
  CN                    &  372.902 &    326.804 &     42.278 \\
  CH$_2$(${}^1$A$_1$)   &  531.275 &    216.344 &      8.168 \\
  NH$_2$                &  493.077 &    258.084 &     10.284 \\
  SH$_2$                &  543.216 &    214.259 &      9.875 \\
  CH2(${}^3$B$_1$)      &  649.215 &    141.029 &      4.072 \\
  SiO                   &  459.809 &    304.986 &     36.431 \\
  N$_2$                 &  472.514 &    434.523 &     41.312 \\
  H$_2$O                &  651.513 &    306.362 &     15.026 \\
  PH$_3$                &  723.682 &    278.878 &      8.265 \\
  CO                    &  725.943 &    320.755 &     34.633 \\
  SO$_2$                &  447.220 &    564.611 &     69.788 \\
  H$_2$O$_2$            &  560.741 &    523.764 &     38.162 \\
  HCO                   &  762.065 &    365.317 &     34.985 \\
  NH$_3$                &  841.268 &    385.909 &     16.307 \\
  HCN                   &  827.211 &    434.722 &     39.711 \\
  SiH$_4$               & 1075.646 &    279.185 &      3.810 \\
  H$_2$CO               & 1075.799 &    451.360 &     33.906 \\
  CO$_2$                & 1026.303 &    536.899 &     59.817 \\
  CH$_3$Cl              & 1242.518 &    386.450 &     22.755 \\
  C$_2$H$_2$            & 1226.379 &    422.963 &     35.796 \\
  CH$_4$                & 1374.066 &    367.050 &     12.345 \\
  N$_2$H$_4$            & 1122.301 &    670.106 &     35.323 \\
  CH$_3$SH              & 1470.091 &    485.920 &     26.279 \\
  CH$_3$OH              & 1546.212 &    566.730 &     29.298 \\
  C$_2$H$_4$            & 1793.679 &    524.465 &     31.427 \\
  \bottomrule
\end{tabular}

\end{table}

\begin{table}[H]
  \centering
  \caption{%
    Ionization potentials (IPs).
    The interpretation of the columns is as in Table~\ref{tab:si:reactionclosed}.
  }
  \begin{tabular}{lrrr}
  \toprule
  Reaction &    HF    &    CCSD    & (T)  \\
  \midrule
  Si                  &  736.815 &     44.959 &      4.710 \\
  B                   &  776.056 &     15.921 &      4.035 \\
  PH$_2$(${}^1$A$_1$) &  899.494 &     42.593 &      5.107 \\
  PH                  &  931.346 &     45.356 &      5.117 \\
  S                   &  891.226 &    100.008 &      4.794 \\
  SH                  &  885.263 &    111.476 &      6.797 \\
  SH2(${}^2$B$_1$)    &  888.222 &    114.845 &      8.155 \\
  P                   &  955.635 &     53.867 &      5.595 \\
  C                   & 1042.056 &     38.083 &      3.153 \\
  Cl$_2$              & 1045.498 &     70.474 &     -1.767 \\
  O$_2$               & 1205.821 &    -25.801 &     -9.885 \\
  H$_2$O              & 1054.632 &    158.840 &     10.092 \\
  ClF                 & 1126.613 &     96.558 &      1.980 \\
  HCl                 & 1110.361 &    117.113 &      7.824 \\
  Cl                  & 1134.933 &    110.247 &      6.075 \\
  OH                  & 1094.944 &    155.435 &      7.858 \\
  O                   & 1160.072 &    146.196 &      5.335 \\
  CO                  & 1256.449 &     97.480 &      2.174 \\
  N                   & 1340.575 &     57.603 &      3.004 \\
  N$_2$(${}^2$S$_g$)  & 1519.395 &      2.990 &    -14.438 \\
  HF                  & 1373.084 &    173.280 &      9.399 \\
  F                   & 1510.014 &    164.236 &      6.611 \\
  \bottomrule
\end{tabular}

\end{table}

\begin{table}[H]
  \centering
  \caption{%
    Electron affinities (EAs).
    The interpretation of the columns is as in Table~\ref{tab:si:reactionclosed}.
  }
  \begin{tabular}{lrrr}
  \toprule
  Reaction &    HF    &    CCSD    & (T)  \\
  \midrule
  P      &   31.078 &    -94.038 &     -7.435 \\
  NH$_2$ &  103.844 &   -158.824 &    -17.433 \\
  PH     &   16.546 &   -104.266 &     -9.326 \\
  PO     &  -75.123 &    -32.123 &     -0.226 \\
  CH     &  -40.900 &    -65.231 &     -8.102 \\
  C      &  -43.505 &    -69.363 &     -7.682 \\
  SiH    &  -72.351 &    -42.481 &     -6.312 \\
  PH$_2$ &   -4.070 &   -106.120 &    -10.365 \\
  Si     &  -81.724 &    -48.047 &     -6.800 \\
  O      &   55.461 &   -179.319 &    -16.165 \\
  S$_2$  &  -78.114 &    -79.944 &     -3.429 \\
  OH     &   25.985 &   -183.877 &    -19.137 \\
  S      &  -85.119 &   -106.677 &     -9.307 \\
  SH     & -102.303 &   -112.532 &    -11.051 \\
  F      & -113.293 &   -199.609 &    -17.460 \\
  CN     & -278.623 &   -101.386 &     -1.919 \\
  \bottomrule
\end{tabular}

\end{table}

\newpage

\section{Further correlation energy statistics}


\begin{table}[H]
  \centering
  \caption{%
    BSIEs of the Hartree--Fock energies with the reference obtained from AV6Z. 
    BSIEs of conventional CCSD valence correlation energy with the reference
    obtained from [56]-extrapolation.
  }
  \label{tab:si:molchannelstat}
  \begin{tabular}{lccccccccccccccc}
%
%
\toprule
&
&
\multicolumn{2}{c}{REc (kJ/mol)} &
&
\multicolumn{2}{c}{REo (kJ/mol)} &
&
\multicolumn{2}{c}{AEs (kJ/mol)} &
&
\multicolumn{2}{c}{IPs (kJ/mol)} &
&
\multicolumn{2}{c}{EAs (kJ/mol)}
\\
    &
    &
max &
rms &&
max &
rms &&
max &
rms &&
max &
rms &&
max &
rms \\
\midrule
HF @ AVTZ    && 6.519 & 2.523 && 15.598 & 3.563 && 10.489 & 3.023 && 2.964 & 1.258 && 2.040 & 0.833 \\
HF @ AVQZ    && 1.898 & 0.691 &&  6.619 & 1.287 &&  3.689 & 0.726 && 0.859 & 0.282 && 0.593 & 0.259 \\
HF @ AV5Z    && 0.625 & 0.187 &&  2.145 & 0.412 &&  1.226 & 0.238 && 0.200 & 0.076 && 0.131 & 0.069 \\
\midrule
CCSD @ AVDZ  &&  44.190 & 14.396 &&   75.092 &  34.859  &&  118.182 & 56.849  && 44.421 & 29.628 && 29.792 & 20.463 \\
CCSD @ AVTZ  &&  19.844 &  6.744 &&   41.541 &  17.175  &&   41.097 & 21.667  && 18.297 & 11.426 && 12.558 &  7.862 \\
CCSD @ AVQZ  &&   9.974 &  2.731 &&   17.478 &   6.550  &&   16.732 &  8.263  &&  7.527 &  4.503 &&  4.295 &  2.889 \\
CCSD @ AV5Z  &&   5.668 &  1.476 &&    8.772 &   3.170  &&    8.165 &  4.045  &&  3.880 &  2.279 &&  2.258 &  1.465 \\
CCSD @ AV6Z  &&   3.269 &  0.852 &&    5.075 &   1.833  &&    4.721 &  2.340  &&  2.245 &  1.319 &&  1.305 &  0.848 \\
\bottomrule

\end{tabular}

\end{table}

\begin{table}[H]
  \centering
  \caption{%
    BSIE of the valence correlation energy of CCSD(T) for the different reactions.
    Reference energy is obtained from [56]-extrapolation. FPa energies as
    described in main text together with (T*) using 12 FNOs per occupied and 
    [34]-extrapolation from MP2 for the correction. FPb energies include (T*) 
    correction using 20 FNOs per occupied and an MP2 estimate obtained from 
    [45]-extrapolation.
  }
  \label{tab:si:totalcorrenergy}
  \begin{tabular}{lccccccccccccccc}
\toprule
Method
&
& \multicolumn{2}{c}{REc (kJ/mol)}
&
& \multicolumn{2}{c}{REo (kJ/mol)}
&
& \multicolumn{2}{c}{AEs (kJ/mol)}
&
& \multicolumn{2}{c}{IPs (kJ/mol)}
&
& \multicolumn{2}{c}{EAs (kJ/mol)} \\
                      &&   max  &  rms  &&   max  &  rms   &&  max   &  rms  &&   max &   rms &&  max  &    rms \\ 
\midrule
$[$23$]$              && 14.007 & 6.077 && 33.924 & 10.270 && 25.533 & 8.229 && 9.000 & 4.317 && 5.766 & 2.993 \\
$[$34$]$              &&  7.173 & 2.254 &&  4.839 &  2.080 &&  4.196 & 1.940 && 1.914 & 0.796 && 1.716 & 0.894 \\
$[$45$]$              &&  1.255 & 0.465 &&  1.998 &  0.713 &&  1.397 & 0.609 && 0.252 & 0.127 && 0.242 & 0.143 \\
FPa + $\Delta$ ps-ppl + (T*) &&  7.714 & 2.639 &&  9.877 &  3.423 && 11.401 & 2.944 && 3.754 & 1.861 && 7.473 & 2.260 \\
FPb + $\Delta$ ps-ppl + (T*) &&  3.565 & 1.144 &&  2.333 &  1.030 &&  4.255 & 0.908 && 1.874 & 0.844 && 4.179 & 1.115 \\
\bottomrule
\end{tabular}

\end{table}

\begin{table}[H]
  \centering
  \caption{%
    BSIE of the CCSD(T) energy for the different reactions. For the reference 
    the Hartree--Fock energy  is obtained using the AV6Z basis set. The CCSD(T)
    valence correlation part is obtained from [56]-extrapolation. Focal-point
    correlation energies as descibed in Table~\ref{tab:si:totalcorrenergy} with 
    Hartree--Fock energies obtained from AVQZ and AV5Z for FPa and FPb, respectively.
  }
  \label{tab:si:molchannelstat}
  \begin{tabular}{lccccccccccccccc}
\toprule
Method
&
& \multicolumn{2}{c}{REc (kJ/mol)}
&
& \multicolumn{2}{c}{REo (kJ/mol)}
&
& \multicolumn{2}{c}{AEs (kJ/mol)}
&
& \multicolumn{2}{c}{IPs (kJ/mol)}
&
& \multicolumn{2}{c}{EAs (kJ/mol)} \\
                      &&   max  &  rms  &&   max  &  rms   &&  max   &  rms   &&   max &   rms &&  max  &    rms \\ 
\midrule 
$[$23$]$              && 20.195 & 8.360 && 49.522 & 13.512 && 36.022 & 10.933 && 7.863 & 3.423 && 4.994 & 2.472 \\
$[$34$]$              &&  8.097 & 2.635 &&  6.065 &  2.478 &&  3.781 &  1.756 && 2.426 & 1.028 && 2.106 & 1.072 \\
$[$45$]$              &&  1.510 & 0.526 &&  2.027 &  0.762 &&  1.271 &  0.489 && 0.331 & 0.178 && 0.315 & 0.175 \\
FPa + $\Delta$ ps-ppl + (T*) &&  8.638 & 3.037 && 16.496 &  4.352 && 11.727 &  3.421 && 3.660 & 1.674 && 7.641 & 2.238 \\
FPb + $\Delta$ ps-ppl + (T*) &&  3.819 & 1.195 &&  3.198 &  1.248 &&  4.316 &  0.927 && 1.882 & 0.819 && 4.185 & 1.115 \\
\bottomrule
\end{tabular}

\end{table}

\end{document}